\documentclass{jfm}

\usepackage{graphicx}
\usepackage{epstopdf, epsfig}

\shorttitle{Optimal kinematics for energy harvesting}
\shortauthor{E. E. Handy-Cardenas, Y. Zhu and K. S. Breuer}

\title{Optimal kinematics for energy harvesting using favorable wake-foil interactions in tandem oscillating hydrofoils}

\author{Eric E. Handy-Cardenas
  \corresp{\email{eric\_handy-cardenas@brown.edu}},
Yuanhang Zhu\footnotemark\footnotetext{Current address: Department of Mechanical Engineering, University of California Riverside, Riverside, CA 92521, USA},
\and Kenneth S. Breuer}

\affiliation{Center for Fluid Mechanics, School of Engineering, Brown University, Providence, RI 02912, USA}

\begin{document}

\maketitle

\begin{abstract}
The energy harvesting performance of a pair of oscillating hydrofoil turbines in tandem configuration is experimentally studied to determine the optimal kinematics of the array. By characterizing interactions between the wake produced by the leading foil and the trailing foil, the kinematic configuration required to maximize array power extraction is determined. This is done by prescribing leading foil kinematics that produce specific wake regimes, identified by the maximum effective angle of attack, $\alpha_{T/4}$, parameter. The kinematics of the trailing foil are allowed to vary significantly from those of the leading foil. The heave and pitch amplitude, inter-foil phase, and foil separation of the trailing foil are varied within each wake regime and the system performance is evaluated. The power extracted by each foil over an oscillation cycle is measured through force and torque measurements. Wake-foil interactions that yield improvements in trailing foil performance are analyzed with time-resolved Particle Image Velocimetry. Constructive and destructive wake-foil interactions are compared, and it was determined that trailing foil performance could be improved by either avoiding interactions with wake vortices or by interacting directly with them. The latter configuration takes advantage of the wake vortex, and does not see power loss during the oscillation cycle. System power from the two foils is maximized when the leading foil is operated at an intermediate $\alpha_{T/4}$ range, and when the trailing foil avoids collisions with wake vortices. This optimal array configuration sees both foils operating with different kinematics compared to the optimal kinematics for a single oscillating foil.
\end{abstract}

\section{Introduction}
As the need for sustainable energy sources continues to rise, more attention has been paid to tidal flow energy due to its high predictability and abundance when compared to other sources like wind
and solar \citep{UihleinMagagna2016,Khare2022}.
Research into hydrokinetic turbine technologies has grown in the past decade \citep{XiaoZhu2014, YoungLaiPlatzer2014} due to their promise as tidal flow energy extraction devices, as well as for the fluid phenomena that govern their behavior. The main types of tidal energy harvesting turbines are horizontal axis rotary turbines (HAT), vertical axis rotary turbines (VAT), and oscillating foil turbines (OFT). Both HATs and VATs have some significant disadvantages such as high rotational speeds that can affect local wildlife and a tendency to biofoul. In addition, HATs create messy wakes that interfere with downstream turbines, a crucial issue for efficient array deployments. 

First proposed by \citet{MckinneyDelaurier1981}, OFTs extract energy using a wing that moves with a coupled heaving and pitching motion in an oscillating manner with respect to an oncoming flow. The motion can be prescribed \citep{Kim2017, KarakasFenercioglu2016}, semi-passive \citep{HeMoGaoetal2022, SuBreuer2019} or fully passive \citep{Oshkai2022, Zhao2023}, depending on how much of the kinematics are directly controlled. The OFT takes advantage of unsteady force generation, where lift, force, and pitching moments are generated during the oscillation cycle. The high angle of attack achieved results in the turbines operating within the dynamic stall regime, with flow separation a prominent feature and strong coherent vortices generated during the oscillation cycle. \citet{KinseyDumas2008} and \citet{Simpson2009} identified that highly loaded cases see strong leading edge vortices (LEVs) formed on the suction side of the foil, resulting in a low-pressure region that induces additional lift on the foil.

Although single foil performance has been well characterized \citep{KinseyDumas2008, Kim2017},  array configurations are an area of significant research interest \citep{KinseyDumas2012, Ribeiro2021, ZhengBai2022, Zhao2023}. An important characteristic of OFT design is the structured wake that develops behind the hydrofoil. In comparison with a rotary turbine, whose wake is characterized by a cylindrical ``sheath'' of blade tip vortices \citep{MassouhDobrev2007}, the wake behind an OFT develops into an unsteady vortex-dominated flow akin to the von K\'{a}rm\'{a}n vortex street \citep{Simpson2009}.
The impact on performance due to the effects of these wakes interacting with downstream turbines is an important consideration for the successful development of efficient turbine arrays. While a HAT array will have downstream turbines always negatively impacted by interactions with an upstream turbine's wake \citep{Kuang2023}, other types of turbines, such as VATs and OFTs, can harness wake-foil interactions to improve array performance.
The similar phenomenon of fish schooling and hydrodynamic interactions in swimmer arrays has been studied extensively \citep{Ramananarivo2016, Newbolt2019, Li2020, Wei2023}.
Inspired by the schooling phenomenon, \citet{Whittlesey2010} proposed a VAT array design using a potential flow model that optimized the turbines' spatial arrangement to increase array performance.

Harnessing wake-foil interactions to improve an array's energy-harvesting performance depends on properly tuning the turbines' kinematic parameters.
In a tandem foil configuration, these parameters are the reduced frequency, $f^*$, the inter-foil phase, $\psi_{1-2}$, and the inter-foil distance, $S_x$ \citep{KinseyDumas2012, KarakasFenercioglu2016, HeYangMoetal2022}.
Tuning these parameters affects the timing of interactions, the magnitude of these effects, and ultimately the performance of both foils in the array.
In an effort to combine the main parameters affecting the timing of wake interactions into a predictive quantity, \citet{KinseyDumas2012} proposed the \textit{global phase parameter}.
Their results only showed a consistent optimal global phase across cases with similar maximum effective angles of attack.
They noted that their global phase lacked information about the wake velocity faced by the downstream foil, which led to different cases presenting a different optimal phase alignment.
\citet{XuXu2017} studied configurations using potential flow theory and observed a different optimal phase to that of Kinsey and Dumas.
\citet{Ribeiro2021} studied tandem arrays computationally and experimentally by separating different cases into wake regimes, a characterization based on the maximum effective angle of attack and the structure of the wake behind the leading foil.
In follow-up work, \citet{RibeiroFranck2023} were able to identify and classify further wake patterns from simulation results of oscillating hydrofoil turbines through a machine learning approach.
\citet{SuPhd2019} introduced a modified global phase parameter that, based on Particle Image Velocimetry (PIV) measurements, incorporated a decelerated wake velocity with respect to the free stream. \citet{Ribeiro2021} similarly modified the global phase parameter from \citet{KinseyDumas2012} and proposed the \textit{wake phase parameter}, given by:
\begin{equation}\label{eq:wakePhase}
    \Phi=2\pi\frac{S_x}{\bar{u}_p}f^*+\psi_{1-2}.
\end{equation}
They showed that optimal phase alignment could be predicted across different tandem foil arrays. This parameter, which includes the average wake velocity within the swept distance directly upstream from the trailing foil, $\bar{u}_p$, to more accurately describe the mean flow speed between the two foils in the array, gives a sense of the alignment of the trailing foil with respect to primary wake structures. \citet{Rival2011} found two types of wake-foil interactions that resulted either in an increase or decrease in the performance of the trailing foil. \citet{Ma2019} also showed how for a semi-passive system, negative vortex interactions had a more detrimental effect on system performance than the performance gained from favorable interactions. \citet{Zhao2023} did an experimental and numerical study of fully passive flapping foils in tandem configuration and found that the trailing foil always became ``locked in'' by the wake generated from the leading foil, regardless of their initial states. They also found that the highest efficiency cases saw the trailing foil heaving with a larger amplitude than that of the leading foil and displayed higher pitching velocities. \citet{Wang2023} studied a semi-passive tandem foil array in which both foils were interconnected and operated with a spacing of $0.2c<S_x<0.33c$. They showed how the closely inter-connected foils could leverage constructive foil-foil interactions to improve their performance, and noted that this decreased separation improved the deployment density of the array.
In very recent work, \citet{Ribeiro2024} proposed a model to predict the efficiency of tandem foil arrays by relating the effective angle of attack and the change in instantaneous power output due to wake vortex interactions through the use of a coefficient of proportionality. They demonstrate that the model is able to predict direct vortex-foil impingement events as well as weak interactions of the trailing foil solely based on simulations of a single foil, although less accuracy is achieved at high angle of attack wake regimes due to the increasing unsteadiness of the wake.

Previous studies have enforced the same kinematics on both foils in tandem arrays. Although this simplifies the parameter space required to seek optimal array configurations, enforcing single-foil optimal kinematics on all foils in the array might not lead to optimal system performance. The primary objective of this paper is to allow for the two foils to have different kinematics, to seek the settings that result in optimal collective energy harvesting performance, and to understand how the different foil kinematics within the array can result in improved wake-foil interactions.
By prescribing the wake regimes identified by \citet{Ribeiro2021} (and further described by \citet{RibeiroFranck2023}) through the kinematics of the leading foil, we evaluate the performance of the trailing foil within each wake regime over a large range of kinematic parameters.
Performance is determined from force and torque measurements obtained through experiments performed in a water flume.
An overview of the performance of all array configurations tested is presented, and a determination of which kinematic configuration of each foil results in the best-performing array is given.
Finally, using the average power-per-cycle extracted from the array and flow measurements from PIV, comparisons are made between optimal and sub-optimal cases to further investigate the effects of wake-foil interactions on the system.

\section{Experimental methods}
\begin{figure}
    \centering
    \includegraphics[width=0.6\textwidth]{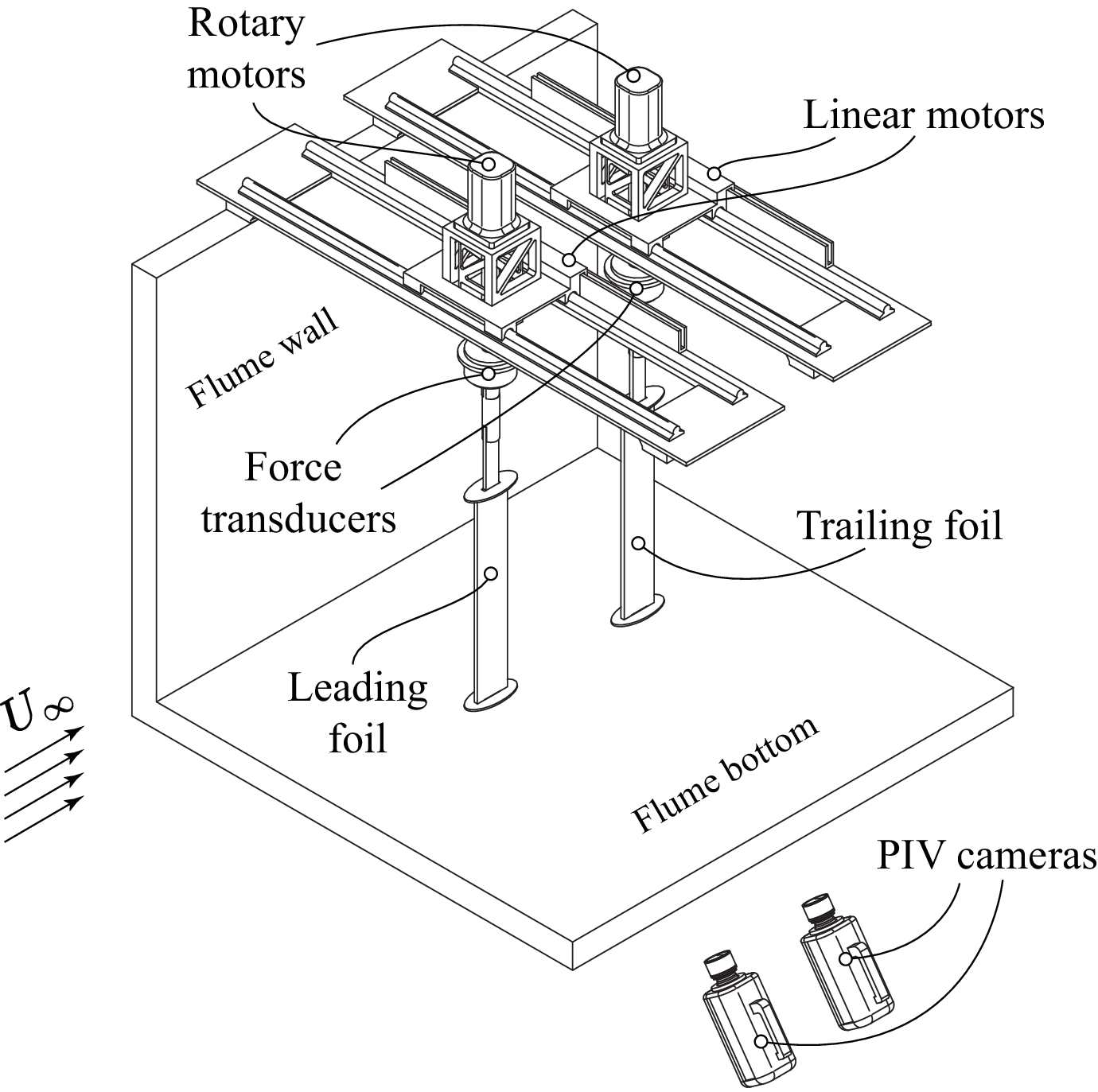}
    \caption{Experimental setup.}
    \label{fig:expSetup}
\end{figure}
Figure~\ref{fig:expSetup} shows a diagram of the experimental setup.
Experiments were performed in an open-channel recirculating flume at Brown University with test section cross-section size 0.8 m $\times$ 0.6 m and 4 m in length.
Two rectangular flat plate hydrofoils were mounted vertically in the flow. 
\citet{Kim2017} tested multiple hydrofoils with different cross-sectional shapes and found that the shape of the foil did not have a significant influence on its energy harvesting performance.  Except where noted, the foils tested measured 0.061 m ($c$) by 0.366 m ($b$) and were fitted with end plates extending $0.75c$ from the hydrofoil edge to their edge, which were used to suppress tip vortex effects.
The heave and pitch motions of the hydrofoils were prescribed by using two gantry traverse systems controlled with servo motors (Parker rotary servo and AeroTech linear servo) mounted on the frame of the flume. The position of each hydrofoil was monitored independently using optical encoders (US Digital).  
Mounted between each traverse and foil, a six-axis force/torque transducer (ATI F/T Delta IP65) recorded force and torque at a sampling rate of 1000 Hz.  The free-stream velocity was set at $0.33$ $\mathrm{m/s}$, measured with an acoustic doppler velocimeter (Nortek Vectrino) positioned upstream, yielding a chord Reynolds number of $\Rey=20,000$. Water temperature was monitored during experiments and did not vary more than $1^\circ$C. 
At each experimental condition, (frequency, heave and pitch amplitude, etc) measurements were taken for 40 oscillation cycles. The first and last five cycles were discarded to guard against any transients, and the middle 30 cycles were subsequently phase-averaged.

Two-component, time-resolved PIV measurements were acquired in an $x$-$y$ plane located at the foil's mid-span.
Two cameras (Photron Fastcam Nova R2, 2048 $\times$ 2048 pixels) were positioned below the test section, side by side, yielding a field of view measuring 0.25 m $\times$ 0.43 m.
The flow field was illuminated using an Nd:YLF laser (Photonics Industries DM30) and image pairs were acquired at 200 Hz (approximately 310 flow fields for each foil oscillation cycle).
The velocity fields were computed using DaVis v10 (LaVision).
Velocity and vorticity fields were then phase-averaged over six oscillation cycles.

\subsection{Oscillating hydrofoil kinematics and energy harvesting performance}
\begin{figure}
    \centering
    \includegraphics[width=\textwidth]{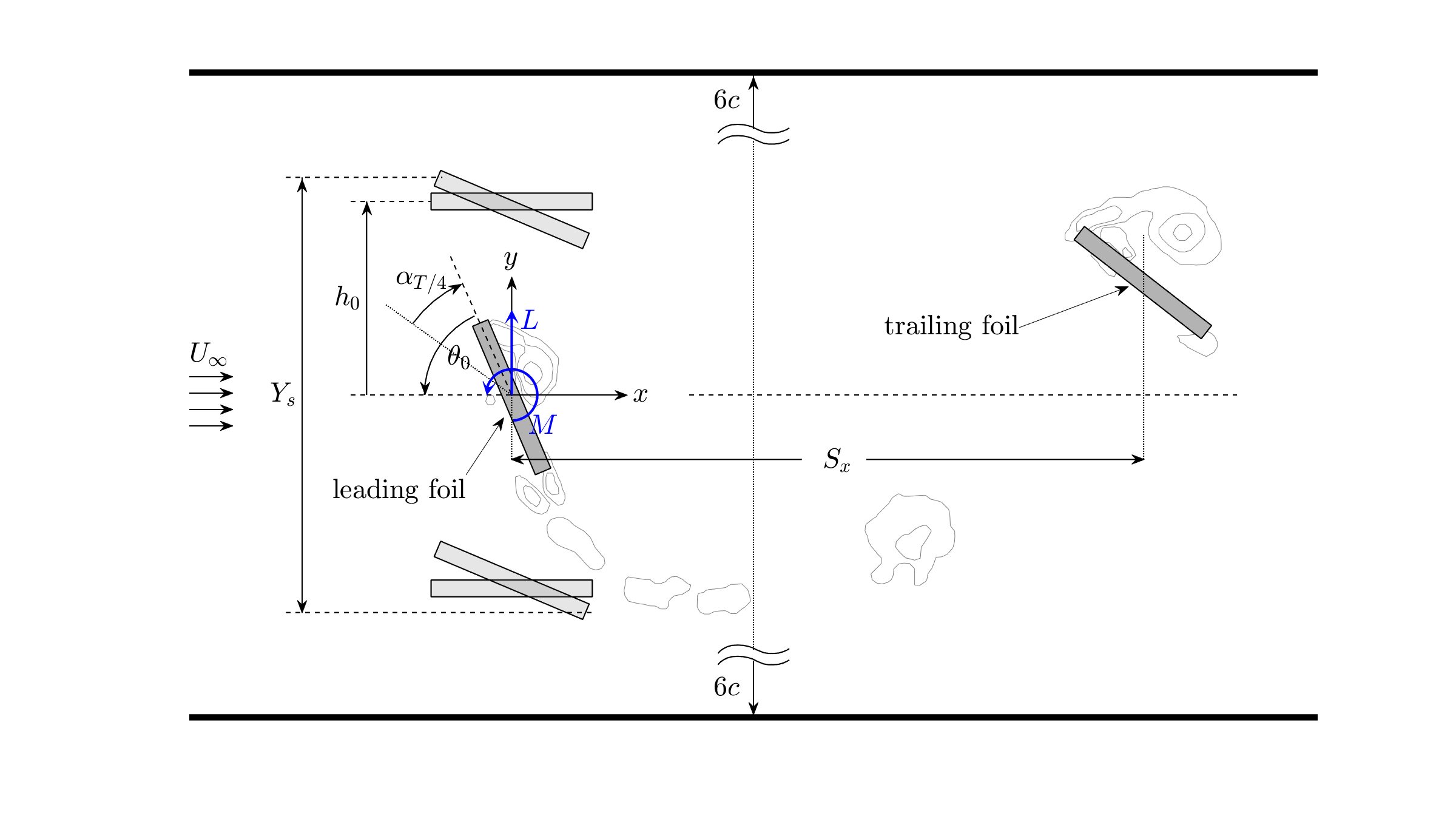}
    \caption{Top view of the measurement space, illustrating the kinematics of a tandem oscillating hydrofoil turbine. Shown are the pitching and heaving amplitudes, $\theta_0$ and $h_0$ respectively, the maximum effective angle of attack $\alpha_{T/4}$, the swept distance $Y_s$, inter-foil separation $S_x$. $L(t)$ and $M(t)$ are the lift force and pitching moment, and $U_{\infty}$ is the free-stream velocity. The distance from the coordinate axis origin to the flume side walls in the transverse direction is 6 chord lengths.}
    \label{fig:kinematicsDiagram}
\end{figure}

The main kinematic parameters governing the operation of the tandem hydrofoil array are shown in figure~\ref{fig:kinematicsDiagram}.
The sinusoidal motion profiles of the hydrofoils are described by 
\begin{equation}\label{eq:heave}
    h(t) = h_0 \sin(2\pi ft - \psi_{1-2}), \ \ \mathrm{and}
\end{equation}
\begin{equation}\label{eq:pitch}
    \theta(t) = \theta_0 \sin(2\pi ft - \pi/2 - \psi_{1-2}), 
\end{equation}
where $h_0$ and $\theta_0$ are the amplitudes of the heave and pitch motions respectively, $f$ is the frequency of oscillation in Hertz, $t$ is time in seconds, and $\psi_{1-2}$ is the phase between the motions of the leading and trailing foils. The phase between heaving and pitching motions was kept constant at $\pi/2$.

We can relate the foil's effective angle of attack, defined as
\begin{equation}\label{eq:aEff}
    \alpha_\mathrm{eff}(t) = \theta(t) - \tan^{-1}(\dot{h}(t)/U),
\end{equation}
to the performance of the foil by using its maximum value, referred to $\alpha_{T/4}$ because it is the effective angle of attack at $t = T/4$, and which is defined as:
\begin{equation}\label{eq:aT4}
    \alpha_{\mathrm{max}}\approx\alpha_{T/4} = \theta_0-\tan^{-1}{(2\pi h_0^*f^*)}, 
\end{equation}
where $h^*_0=h_0/c$ is the non-dimensional heave and $f^*=fc/U_\infty$ the reduced frequency.
Previous studies \citep{KinseyDumas2008, Simpson2009, Kim2017, Ribeiro2021, RibeiroFranck2023} have shown that $\alpha_{T/4}$ effectively parametrizes the energy harvesting performance of an oscillating foil, (see figure~\ref{fig:aT4vsEffV2}).

The energy harvesting performance of the hydrofoil is obtained from the sum of the power extracted due to the lift force, $L$, and the pitching moment, $M$:
\begin{equation}\label{eq:power}
    P(t)=L(t)\dot{h}(t)+M(t)\dot{\theta}(t).
\end{equation}
Energy harvesting (or Betz) efficiency, $\eta$, is defined as the ratio between the power extracted per cycle by the foil and the power available in the oncoming flow:
\begin{equation}\label{eq:efficiency}
    \eta=\overline{P(t)}/0.5\rho U_{\infty}^{3}A_s , 
\end{equation}
where $\rho$ is the fluid density, $c$ is the chord length, $b$ the span, and $A_s$ the swept area given by $A_s=bY_s$. $\overline{P(t)}$ is the cycle-averaged power.

\subsection{Blockage correction}
Hydrofoil performance is affected by blockage effects due to experiments being performed in the flume which has side walls. We account for these effects using a blockage correction model from \citet{Maskell1965}, as implemented by \citet{RossPolagye2020}. This model couples axial momentum theory with assumptions about the wake behind a highly loaded turbine.  Information on the bypass flow is required to make use of this correction, which, for this study, is obtained using the open-channel blockage correction model developed by \citet{Houlsby2008}, as implemented by \citet{Ribeiro2021}.

\subsection{Kinematic parameter selection}

Several series of experiments were conducted.  In order to validate our system, we first characterized single foil performance, measuring forces and torques over a range of frequencies, heave, and pitch amplitudes (Table~\ref{tab:singleFoilParameters}). In addition, two foil sizes were tested with chord measuring 0.061 m and 0.0762 m.  The plates had thicknesses measuring 0.00635 m and 0.0079 m respectively, and had the same aspect ratio. The freestream velocity was adjusted to preserve a constant Reynolds number (20,000). Thus the only parameter changing was the ratio of chord to flume width, $c/W$.

Secondly, we conducted a parameter sweep study of the energy harvesting performance (i.e. force/torque, no PIV) of tandem foils.  In this series of measurements (Table~\ref{tab:tandemFoilParameters}) we chose three leading foil operating points representing the shear layer, LEV and LEV+TEV flow regimes \citep{Ribeiro2021,RibeiroFranck2023}.  
For each of the three leading foil operating kinematics, a full parameter sweep of the \textit{trailing foil} was performed by varying $h_\mathrm{tr}$, $\theta_\mathrm{tr}$, $\psi_{1-2}$, and $S_x$, and the performance of the full system was quantified by calculating the power extracted by the tandem foil combination.
A total of 1,323 combinations of kinematic parameters were tested.

Lastly, PIV was carried out for 16 parameter combinations - chosen to provide detailed flow information on the constructive and destructive wake-foil interactions identified during the full parameter sweep. The parameters for these cases are shown in  Table~\ref{tab:PIVParameters}.

\begin{table}
    \begin{center}
\def~{\hphantom{0}}
    \begin{tabular}{c|c|c|c}
    $f^*$ & $c/W$ & $h_{\mathrm{0}}$ & $\theta_{\mathrm{0}}$ \\ [6pt]
    $0.10$ & 0.1, 0.07 & $0.5c$ to $1.5c$ & $40^\circ$ to $80^\circ$ \\
    $0.12$ & 0.1, 0.07 & $0.5c$ to $1.5c$ & $40^\circ$ to $80^\circ$ \\
    $0.15$ & 0.1, 0.07 & $0.5c$ to $1.5c$ & $40^\circ$ to $80^\circ$ \\
    \end{tabular}
    \caption{Single foil experimental parameters explored in this study. Reduced frequency ($f^*$), heaving amplitude ($h_0$), and pitching amplitude ($\theta_0$) were varied for foils with two different chord-to-test section ratios, $c/W$. The combination of the selected parameters yielded values of $\alpha_{T/4}$ ranging from $0.025$ rad to $0.95$ rad.}
    \label{tab:singleFoilParameters}
    \end{center}
\end{table}

\begin{table}
    \begin{center}
\def~{\hphantom{0}}
    \begin{tabular}{c|c|c|c|c|c|c|c|c}
    Wake regime & $\alpha_{T/4}^\mathrm{le}$ & $f^*$ & $S_x$ & $h_{\mathrm{le}}$ & $\theta_{\mathrm{le}}$ & $h_{\mathrm{tr}}$ & $\theta_{\mathrm{tr}}$ & $\psi_{1-2}$ \\ [6pt]
    Shear layer & 0.16 & 0.12 & 4$c$ & 0.8$c$ & $40^\circ$ & $\{0.6c$, $0.8c$,  & $65^\circ$, $70^\circ$, $75^\circ$    & $\{-180^\circ$, $-110^\circ$, \\
    Shear layer & 0.16 & 0.12 & 6$c$ & 0.8$c$ & $40^\circ$ & $1.0c$, $1.2c$,    & $65^\circ$, $75^\circ$                & $-52^\circ$, $0^\circ$, \\
    Shear layer & 0.18 & 0.11 & 4$c$ & 1.2$c$ & $50^\circ$ & $1.4c$, $1.6c$,    & $65^\circ$, $75^\circ$                & $52^\circ$, $110^\circ$, \\
    LEV         & 0.33 & 0.12 & 4$c$ & 0.8$c$ & $50^\circ$ & $1.8c\}$,          & $65^\circ$, $70^\circ$, $75^\circ$    & $180^\circ\}$ \\
    LEV         & 0.33 & 0.12 & 6$c$ & 0.8$c$ & $50^\circ$ &                    & $65^\circ$, $75^\circ$                &  \\
    LEV         & 0.35 & 0.11 & 4$c$ & 1.2$c$ & $60^\circ$ &                    & $65^\circ$, $75^\circ$                &  \\
    LEV+TEV     & 0.68 & 0.12 & 4$c$ & 0.8$c$ & $70^\circ$ &                    & $65^\circ$, $70^\circ$, $75^\circ$    &  \\
    LEV+TEV     & 0.68 & 0.12 & 6$c$ & 0.8$c$ & $70^\circ$ &                    & $65^\circ$, $75^\circ$                &  \\
    LEV+TEV     & 0.70 & 0.11 & 4$c$ & 1.2$c$ & $80^\circ$ &                    & $65^\circ$, $75^\circ$                &  \\
    \end{tabular}
    \caption{Tandem foil experimental parameters explored in this study. The leading foil's heaving ($h_\mathrm{le}$) and pitching ($\theta_\mathrm{le}$) amplitudes, as well as the reduced frequency ($f^*$) were set to obtain a desired $\alpha_{T/4}^\mathrm{le}$ within each wake regime. Different combinations of inter-foil separation ($S_x$), inter-foil phase ($\psi_{1-2}$), and heaving ($h_\mathrm{tr}$) and pitching ($\theta_\mathrm{tr}$) amplitudes of the trailing foil were then tested.}
    \label{tab:tandemFoilParameters}
    \end{center}
\end{table}

\begin{table}
    \begin{center}
\def~{\hphantom{0}}
    \begin{tabular}{c|c|c|c|c|c|c|c|c}
    Wake regime & $\alpha_{T/4}^\mathrm{le}$ & $f^*$ & $S_x$ & $h_{\mathrm{le}}$ & $\theta_{\mathrm{le}}$ & $h_{\mathrm{tr}}$ & $\theta_{\mathrm{tr}}$ & $\psi_{1-2}$ \\ [6pt]
    Shear layer & 0.16 & 0.12 & 6$c$ & 0.8$c$ & $40^\circ$ & $0.7c$ & $75^\circ$ & $60^\circ$, $-120^\circ$ \\
    Shear layer & 0.16 & 0.12 & 4$c$ & 0.8$c$ & $40^\circ$ & $0.8c$, $1.2c$ & $75^\circ$ & $0^\circ$ \\
            LEV & 0.33 & 0.12 & 6$c$ & 0.8$c$ & $50^\circ$ & $0.7c$ & $75^\circ$ & $60^\circ$, $-120^\circ$ \\
            LEV & 0.33 & 0.12 & 4$c$ & 0.8$c$ & $50^\circ$ & $0.8c$ & $75^\circ$ & $60^\circ$ \\
            LEV & 0.33 & 0.12 & 4$c$ & 0.8$c$ & $50^\circ$ & $1.4c$ & $75^\circ$ & $-60^\circ$ \\
        LEV+TEV & 0.68 & 0.12 & 6$c$ & 0.8$c$ & $70^\circ$ & $0.7c$ & $75^\circ$ & $60^\circ$, $-110^\circ$ \\
        LEV+TEV & 0.68 & 0.12 & 6$c$ & 0.8$c$ & $70^\circ$ & $0.8c$ & $75^\circ$ & $60^\circ$, $-110^\circ$ \\
        LEV+TEV & 0.68 & 0.12 & 4$c$ & 0.8$c$ & $70^\circ$ & $0.8c$ & $75^\circ$ & $60^\circ$, $180^\circ$ \\
        LEV+TEV & 0.68 & 0.12 & 4$c$ & 0.8$c$ & $70^\circ$ & $1.4c$ & $75^\circ$ & $-110^\circ$, $180^\circ$ \\
    \end{tabular}
    \caption{Tandem foil PIV experiments performed in this study. Wake regimes were prescribed by obtaining a desired $\alpha_{T/4}^\mathrm{le}$ through the choice of the leading foil's reduced frequency ($f^*$), heaving ($h_\mathrm{le}$) and pitching ($\theta_\mathrm{le}$) amplitudes. PIV was then performed on selected combinations of trailing foil heaving ($h_\mathrm{tr}$) and pitching ($\theta_\mathrm{tr}$) amplitudes, inter-foil separation ($S_x$) and inter-foil phase ($\psi_{1-2}$).}
    \label{tab:PIVParameters}
    \end{center}
\end{table}

\section{Results and Discussion}

\subsection{Single foil performance and blockage correction}

\begin{figure}
    \centerline{\includegraphics[width=\textwidth]{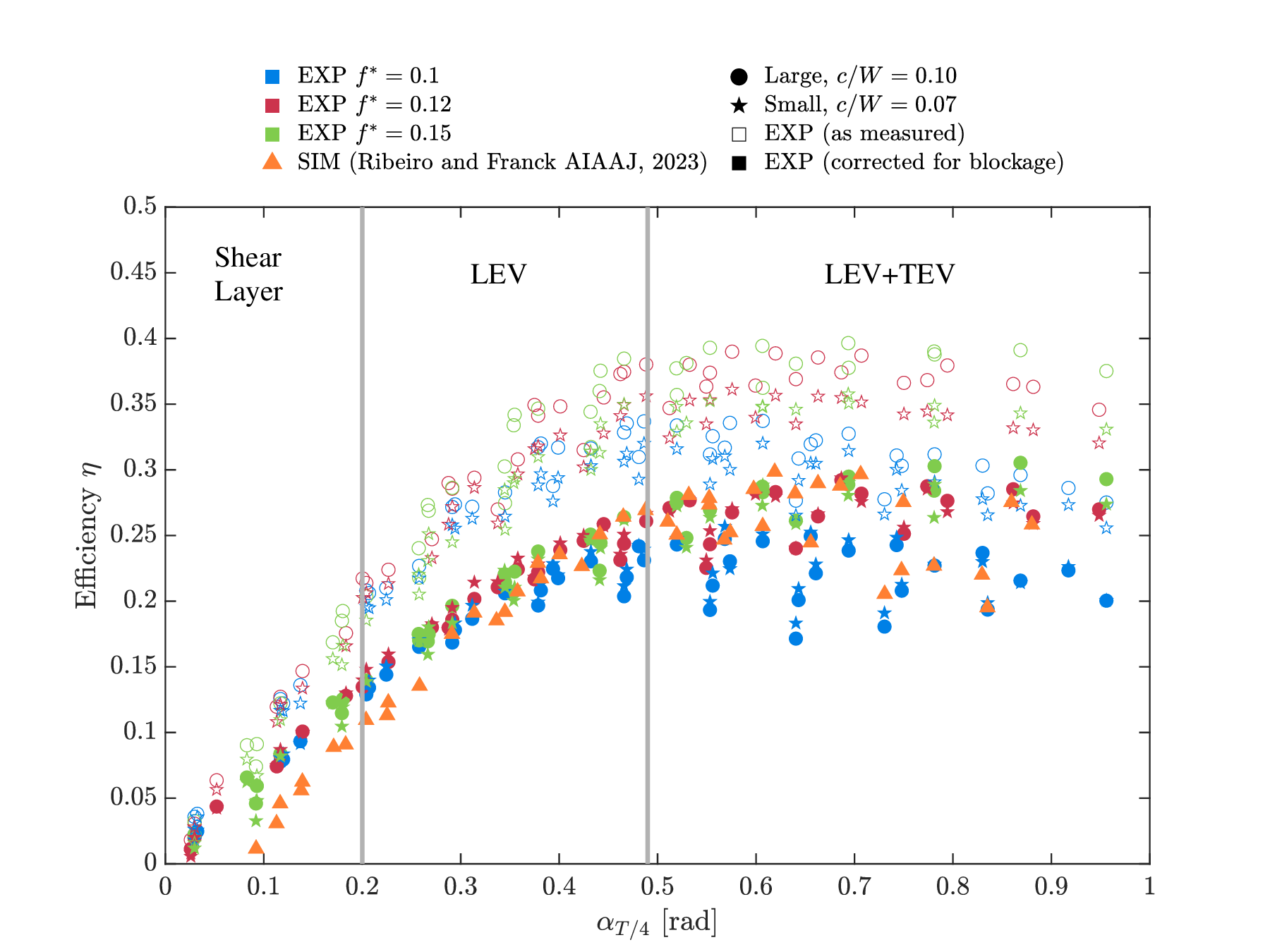}}
    \caption{Energy harvesting efficiency of a single foil as a function of its $\alpha_{T/4}$ value. Blue, red and green markers indicate different reduced frequencies. Orange $\blacktriangle$ markers indicate simulation results from \citet{RibeiroFranck2023}. $\circ$ and $\star$ markers correspond to the efficiency calculated from two differently scaled hydrofoils, where $c/W$ is the ratio of the foil's chord to the flume width. Open markers are measured values and filled-in markers are blockage-corrected values \citep{Maskell1965,RossPolagye2020}.}
    \label{fig:aT4vsEffV2}
\end{figure}

Figure~\ref{fig:aT4vsEffV2} shows the Betz efficiency, $\eta$, of a single foil, as a function of $\alpha_{T/4}$ measured over a range of frequencies, pitch and heave amplitudes, and blockage, $c/W$ (Table~\ref{tab:singleFoilParameters}).
Several features should be noted.  Firstly we see that the efficiency of the larger foil, ($c/W = 0.10$, open $\circ$ markers) is higher than that of the smaller foil ($c/W = 0.07$, open $\star$ markers) and that the performance of both foils decreases and almost perfectly collapses onto a single curve once the blockage correction is applied (filled $\bullet$ and $\star$ markers).
With the blockage correction applied, both sets of experimental data agree very well with the computations of \citet{RibeiroFranck2023} (filled orange $\blacktriangle$ markers), which were performed for an elliptical foil in an infinite (unconstrained) domain.

Secondly, once the blockage correction is applied, (filled symbols) there is excellent scaling of $\eta$ within the shear layer and LEV regimes ($\alpha_{T/4} < 0.5$), while in the LEV+TEV regime ($\alpha_{T/4} > 0.5$) we see the efficiency spread out with the lower frequencies (blue symbols) exhibiting a lower efficiency than the higher frequencies (red and green symbols). This behavior agrees with the computations of \citet{RibeiroFranck2023} (orange symbols), who also observed this bifurcation, and remarked that the split in the $\alpha_{T/4}$ scaling in the LEV+TEV regime divided according to the frequency, with the upper branch corresponding to cases with $f^*=0.15$ and $f^*=0.12$, while the lower branch corresponded to cases with a lower frequency, $f^*=0.1$.

\subsection{Wake structure and vortex trajectories behind a single hydrofoil}
\label{sec:wakeStructure}
\begin{figure}
    \centering
    \includegraphics[width=\textwidth]{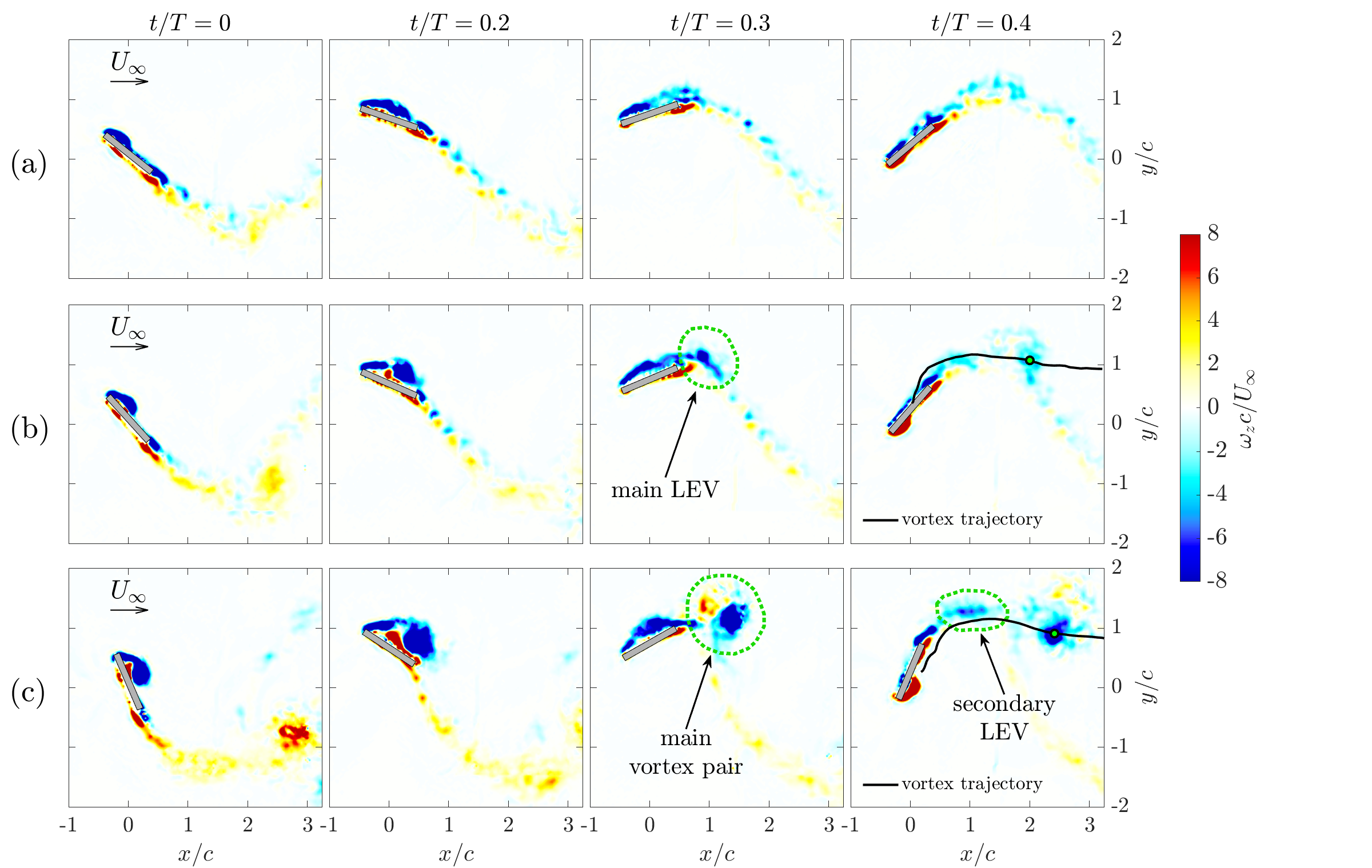}
    \caption{Countours of vorticity for each wake regime at different times within an oscillation cycle. In all three cases the leading foil's reduced frequency is $f^*=0.12$ and its heaving amplitude is $h_\mathrm{le}=0.8c$. (a) Shear layer regime, $\alpha_{T/4}^\mathrm{le}=0.16$, $\theta_\mathrm{le}=40^\circ$. (b) LEV regime, $\alpha_{T/4}^\mathrm{le}=0.33$, $\theta_\mathrm{le}=50^\circ$. (c) LEV+TEV regime, $\alpha_{T/4}^\mathrm{le}=0.68$, $\theta_\mathrm{le}=70^\circ$. The main vortices in the LEV and LEV+TEV regimes are highlighted in the last snapshot, as well as the trajectory of the primary vortex.}
    \label{fig:wakeStructures}
\end{figure}

The three wake regimes identified by \citet{Ribeiro2021} are confirmed by PIV measurements behind a single foil, and are shown in figure~\ref{fig:wakeStructures} at four times during the cycle.
At the lowest angles of attack, over a range of $0<\alpha_{T/4}\leq0.2$, the ``shear layer regime'' is characterized by a shear layer in the wake with only weak vortex formation. At intermediate angles of attack, we can identify the ``leading edge vortex regime'' (LEV regime) at values of $0.2<\alpha_{T/4}\leq0.49$.
\citet{Lee2022} showed how the primary primary LEV shed by the foil in this regime is advected downstream in a mostly straight downstream direction from its detachment location. This behavior is confirmed by the vortex path shown in the last panel of figure~\ref{fig:wakeStructures}b.
The highest range of $\alpha_{T/4}$ denotes the ``leading edge vortex + trailing edge vortex regime'' (LEV+TEV) which sees a vortex pair shed by the foil every stroke.
The primary LEV is stronger and bigger than in the other wake regimes and is accompanied by a small counter-rotating TEV.
\citet{Lee2022} also showed that the orbiting interaction between these two vortices in this regime results in a curved trajectory as the vortex pair travels downstream 
(figure~\ref{fig:wakeStructures}c, $t/T = 0.4)$.

\begin{figure}
    \centering
    \includegraphics[width=\textwidth]{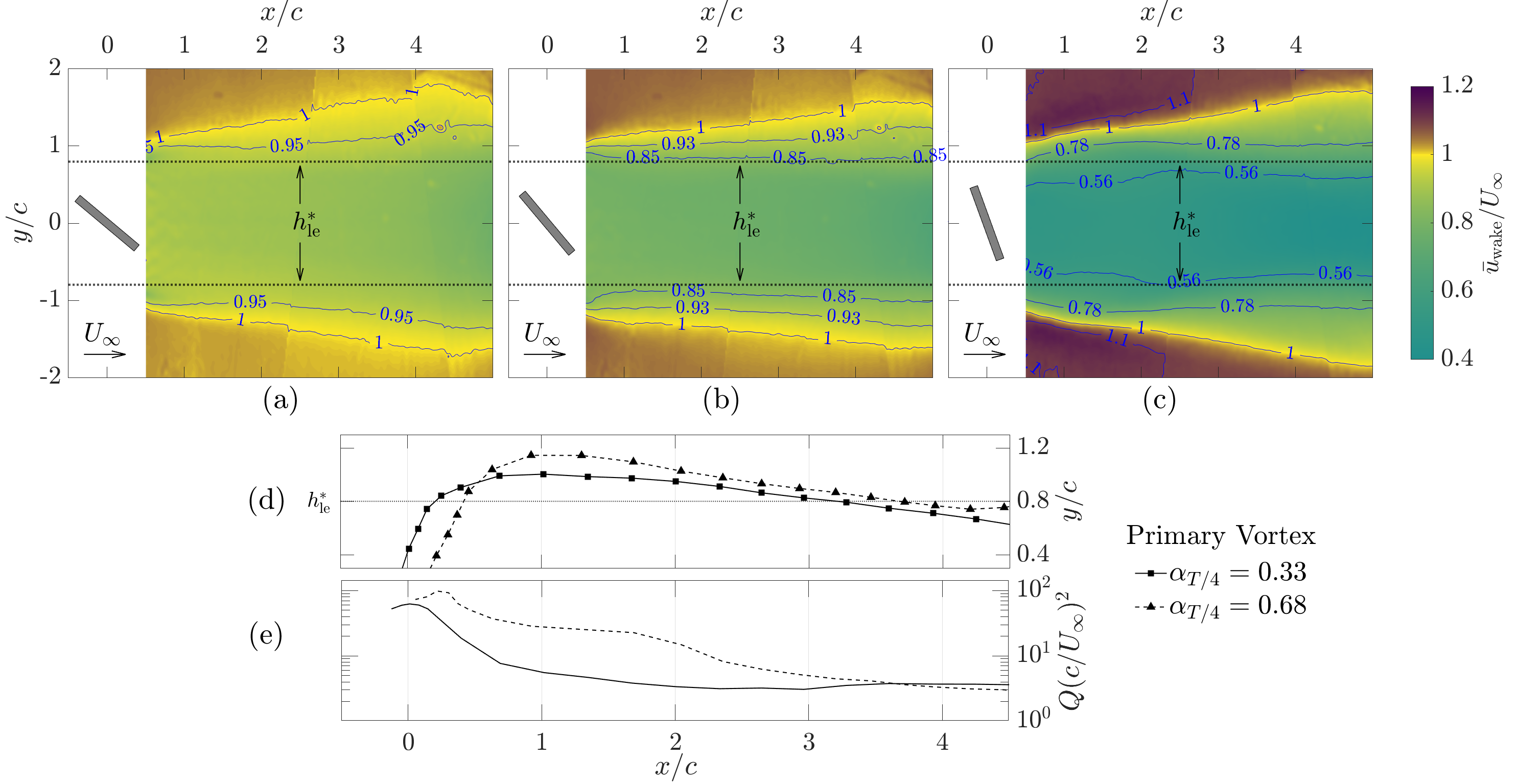}
    \caption{Average wake velocity calculated from PIV measurements for each wake regime, obtained from time-averaging the instantaneous wake velocity over 10 oscillation cycles.}
    \label{fig:wakeVelocityAndVortices}
\end{figure}
Figure~\ref{fig:wakeVelocityAndVortices} shows the average wake velocity behind the leading foil for each wake regime. These results, which align closely with those reported by \citet{Ribeiro2021}, show how the wake velocity within the confines of the leading foil's heaving amplitude decreases at higher values of $\alpha_{T/4}$. The leading foil's heaving amplitude is fixed at $h_\mathrm{le}=0.8c$ in all wake regimes, setting the shedding location of the primary LEV. Although the trajectory of the primary LEV is slightly different in both wake regimes, the vortex remains within the same $y/c$ region as it nears the position of the trailing foil. Due to the presence of the strong vortices, this region in the wake is characterized by high turbulent kinetic energy, as shown by \citet{Ribeiro2021}, which defines a sort of boundary of the pure deficit region. Figures~\ref{fig:wakeVelocityAndVortices}b and \ref{fig:wakeVelocityAndVortices}c show the main vortex trajectory and $Q$-value, where $Q$ is the second invariant of the velocity gradient tensor:
\begin{equation}
 Q=\frac{1}{2}(\|\boldsymbol{\Omega}\|^2-\|\boldsymbol{S}\|^2),
\label{eq:Q-def}   
\end{equation}
for the two wake regimes with prominent vortex structures. It can be inferred from figure~\ref{fig:wakeVelocityAndVortices} that the $h_\mathrm{le}$ indirectly affects $h_\mathrm{tr}$, as it defines the region of decelerated flow that the trailing foil will have to operate in.

\subsection{Array performance}\label{sec:systemPerformance}
\begin{figure}
    \centering
    \includegraphics[width=0.65\textwidth]{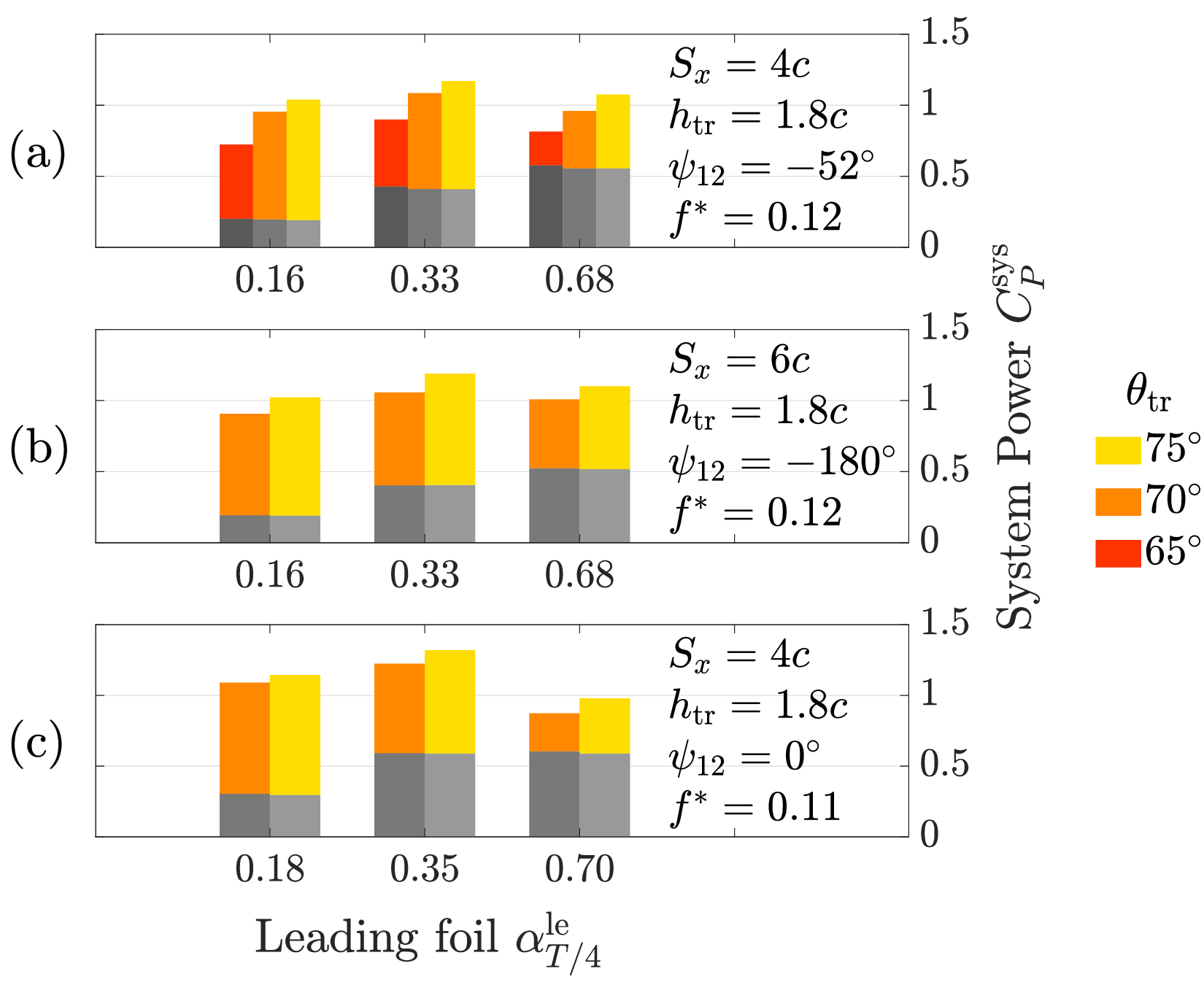}
    \caption{Power extracted by the system as a function of the trailing foil pitch amplitude $\theta_{\mathrm{tr}}$. The grey bars are the power extracted by the leading foil, while the colored bars are the power from the trailing foil.}
    \label{fig:stackedPowerPitch}
\end{figure}
\begin{figure}
    \centering
    \includegraphics[width=0.72\textwidth]{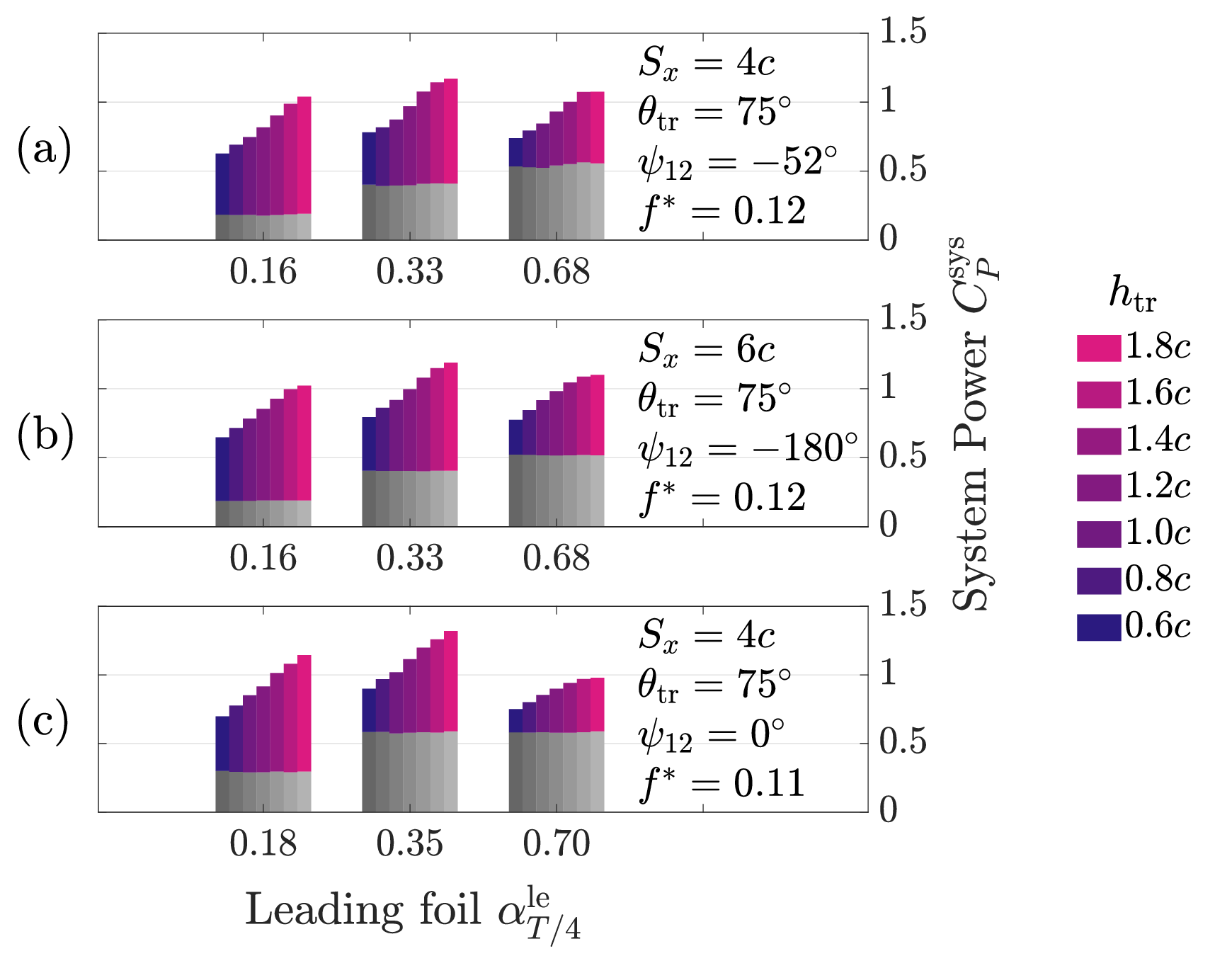}
    \caption{Power extracted by the system as a function of the trailing foil heave amplitude $h_{\mathrm{tr}}$. The grey bars are the power extracted by the leading foil, while the colored bars are the power from the trailing foil.}
    \label{fig:stackedPowerHeave}
\end{figure}
\begin{figure}
    \centering
    \includegraphics[width=0.72\textwidth]{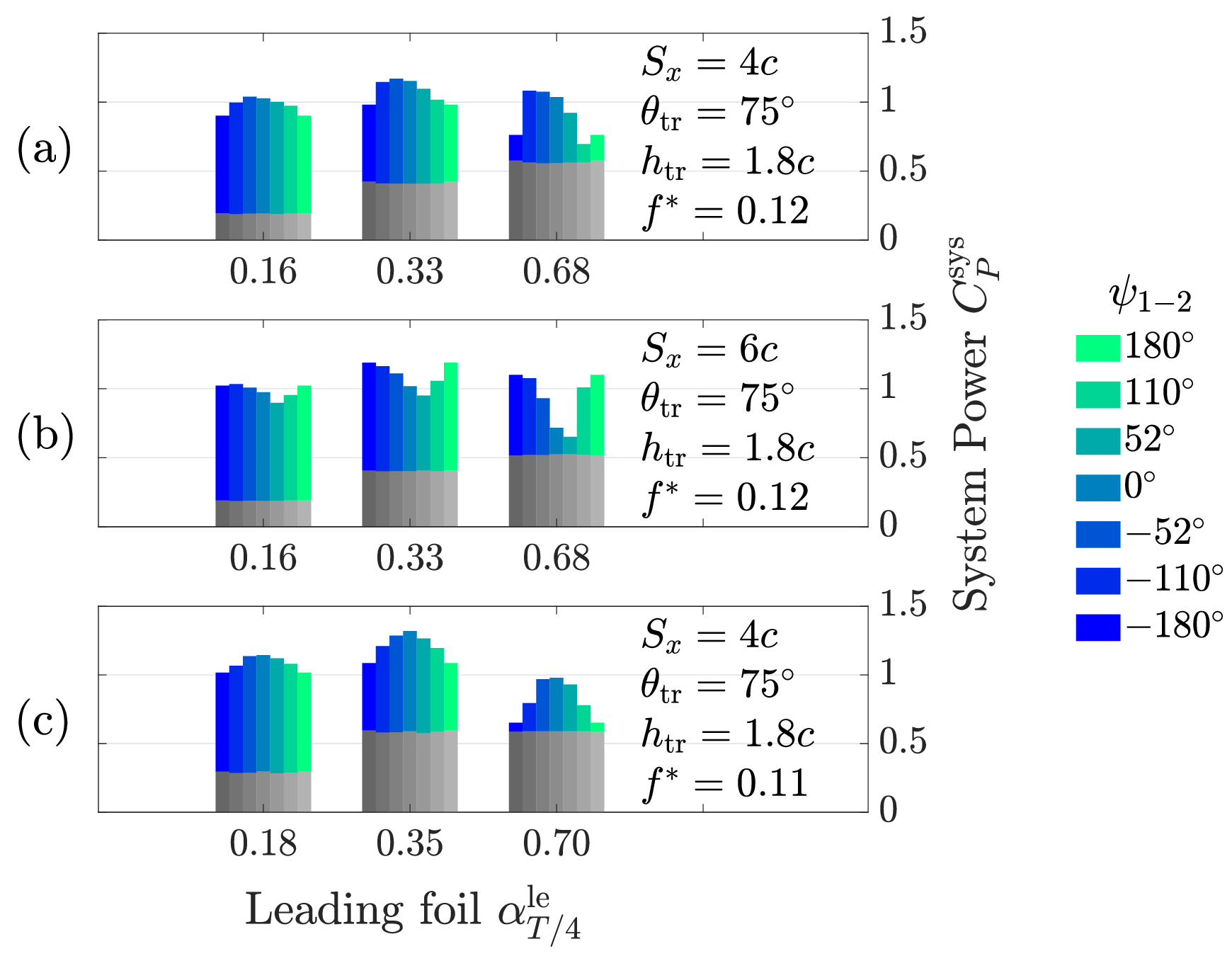}
    \caption{Power extracted by the system as a function of inter-foil phase $\psi_{12}$. The grey bars are the power extracted by the leading foil, while the colored bars are the power from the trailing foil.}
    \label{fig:stackedPowerPhase}
\end{figure}

Figures~\ref{fig:stackedPowerPitch} to \ref{fig:stackedPowerPhase} present a cross-sectional overview of the system power extraction results of the two-foil tandem array.
The total power coefficient of the system is given by:
\begin{equation}\label{eq:systemPower}
    C_P^\mathrm{sys}=\frac{\overline{P_{\mathrm{le}}(t)}+\overline{P_\mathrm{tr}(t)}}{0.5 \rho U_\infty^3 cb},
\end{equation}
and each figure shows the variation of the system power as a function of different parameter combinations for the array. Within each figure, there are three sub-figures, each representing a different sampling of the trailing foil parameters. Each sub-figure contains three groups of power coefficients representing the Shear Layer, LEV, and LEV+TEV regimes, defined by three values of $\alpha_{T/4}$.
Lastly, in each group of $C_P$ plots, a bar-graph ``stack'' is shown in which a specific parameter is systematically varied: $\theta_\mathrm{tr}$, $h_\mathrm{tr}$ and $\psi_{1-2}$ (figures~\ref{fig:stackedPowerPitch}, \ref{fig:stackedPowerHeave} and \ref{fig:stackedPowerPhase} respectively).
Each bar-graph stack shows the leading foil $C_P^\mathrm{le}$ in grey (which is more-or-less constant within each group, since the leading foil kinematics are fixed for that group); the trailing foil performance, $C_P^\mathrm{tr}$, is shown in color.

Results from \citet{KinseyDumas2012} and \citet{Xu2016} identified that $S_x$ mainly affects the timing of wake-foil interactions.
And while other parameters like $f^*$ and $\psi_{1-2}$ also affect directly these interactions, $S_x$ further influences the energy extracted by the downstream foil due to the wake velocity deficit.
Subplots (a) and (b) in figures~\ref{fig:stackedPowerPitch} to \ref{fig:stackedPowerPhase} correspond to cases that have the same $\alpha_{T/4}^\mathrm{le}$ values but different inter-foil separations ($S_x=4c$ in (a) and $6c$ in (b)), therefore the only parameter changing is $S_x$ (at optimal phasing, $\psi_{1-2}$). We observe that the magnitude of $C_P^\mathrm{sys}$ does not vary significantly due to $S_x$, which we can explain with figures~\ref{fig:wakeVelocityAndVortices}a to c, where it is observed that the wake velocity does not recover very much between these two separations in all wake regimes, thus the trailing foil faces approximately the same decreased upstream flow velocity for both separations tested. Considering the negligible variation of wake velocity for the tested $S_x$, it might be desirable to space the foils closely to increase the number of turbines deployed in a given area, i.e. the power density of the array. As shown by figure~\ref{fig:wakeVelocityAndVortices}e, decreasing $S_x$ would ensure that vortices being shed from the leading foil are stronger and more coherent, and therefore could be harnessed by the trailing foil more effectively. However, this also means that strong negative vortex-foil interactions could also occur, making optimizing the array kinematics all the more important. It is important to note that decreased leading foil performance can result from spacing the foils closer together, as observed by \citet{Xu2016}, who noted that the blockage effect of the trailing foil on the leading foil becomes more important at these close separations. This detrimental effect was also shown by \citet{Zhao2023} in their passive system who found that as $S_x$ is decreased near $2c$, the trailing foil's influence on the leading foil can lead to decreased leading foil performance.

Figure~\ref{fig:stackedPowerPitch} demonstrates that increases in the trailing foil pitch amplitude, $\theta_\mathrm{tr}$,  correlate with increased trailing foil performance.
Although the physical frequency is fixed for both foils in the array, the effective non-dimesional frequency, $f^* =fc/U$, of the trailing foil will increase due to the diminished velocity faced by the trailing foil.respect to the leading.
In their single foil experiments \citet{Kim2017} showed that the optimal pitch amplitude rises as $f^*$ rises, and this same trend was observed in the numerical simulations of \citet{KinseyDumas2008}.
This effect can be explained by observing that the relative speed between the oncoming flow and the trailing foil is lower (leading to the increased $f^*$), which results in a lower maximum effective angle of attack ($\alpha^\mathrm{tr}_{T/4}$). This can be compensated for by a shift of the optimal $\theta_\mathrm{tr}$ towards higher amplitudes, perhaps even beyond the maximum value tested ($75^\circ$) in this sequence of experiments.

Figure~\ref{fig:stackedPowerHeave} presents the system performance dependency on the trailing foil heave amplitude and demonstrates that the performance of the trailing foil rises with increasing $h_\mathrm{tr}$ in all three wake regimes.
This was also observed for a single foil by \citet{KinseyDumas2008}, and although their work focused on the efficiency of the foil (which is scaled by the swept area), they remarked that by increasing the foil's $h_0$ a higher $C_P$ (which is scaled by the foil area) was obtained despite a lower efficiency. Additionally, by allowing $h_\mathrm{tr}$ to be much larger than $h_\mathrm{le}$, the trailing foil can access higher momentum fluid and thus improve its $C_P$.
Note how figure~\ref{fig:wakeVelocityAndVortices} shows an increase in the amplitude of the bypass flow behind the leading foil, most notably in the LEV+TEV regime where it reaches up to $\bar{u}_\mathrm{wake}/U_\infty\sim1.1$.
As with the pitch amplitude discussion earlier, it is possible that our test matrix did not explore sufficiently large values of $h_\mathrm{tr}/c$, although the incremental improvement does seem to be flattening out.  Interestingly, \citet{Zhao2023} also found that the best performance of the trailing foil was achieved for higher heaving and pitching amplitudes at inter-foil separations ranging from $S_x=1c$ to $6c$, even though they had a fully passive system.
Notably, the local $C_P^\mathrm{tr}$ maximum is only reached at the LEV+TEV wake regime at $h_\mathrm{tr}=1.6c$, after which it plateaus.
One possible reason for this is that the more unsteady wake of the LEV+TEV regime limits the power extraction of the trailing foil at very high $h_\mathrm{tr}$ due to the presence of strong LEVs in this region.

The performance variations of the trailing foil with respect to changing the inter-foil phase, $\psi_{1-2}$, are presented in figure~\ref{fig:stackedPowerPhase}.
Many studies \citep{Ashraf2011,KinseyDumas2012,KarakasFenercioglu2017,Ribeiro2021} have explored the effect of this critical parameter on the performance of the array.
For foils sharing the same $f^*$, $\theta_0$, and $h_0$, \citet{Ribeiro2021} evaluated the tandem array's performance for a large range of $\psi_{1-2}$ and observed the same harmonic trends that figure~\ref{fig:stackedPowerPhase} demonstrates.
The effect of $\psi_{1-2}$ becomes more pronounced as $\alpha_{T/4}^\mathrm{le}$ is increased due to the stronger wake features, leading to variations of different magnitude in $C_P^\mathrm{tr}$ for each wake regime ($\pm 0.05$ in the shear layer regime and $\pm 0.25$ in the LEV+TEV regime).
The timing of wake-foil interactions, dictated mainly by $\psi_{1-2}$, will lead to optimal or sub-optimal parameter combinations, which can be more easily identified by using the \textit{wake phase parameter}, $\Phi$, from \citet{Ribeiro2021} (see Eqn. \ref{eq:wakePhase}).
\begin{figure}
    \centering
    \includegraphics[width=\textwidth]{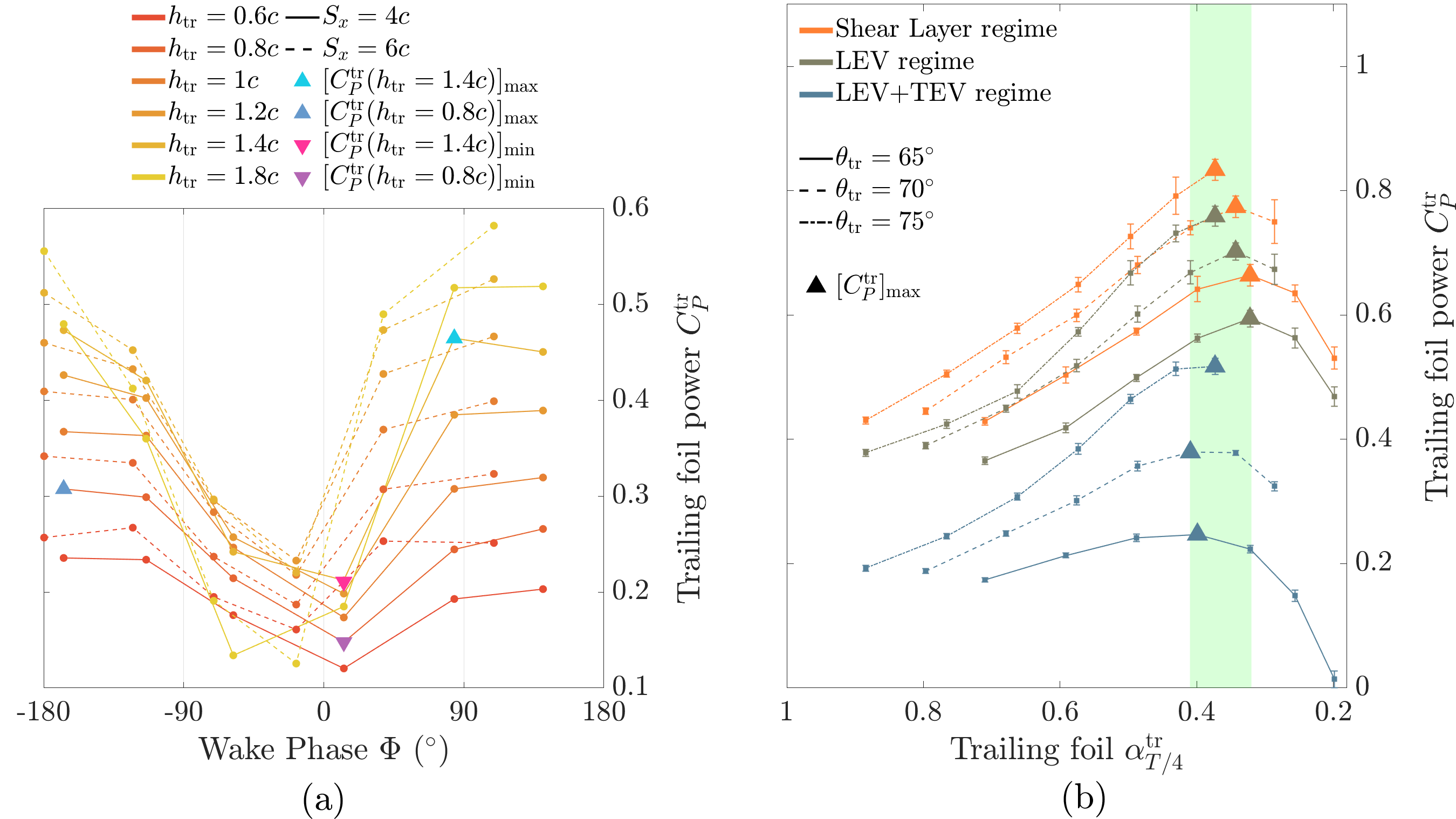}
    \caption{(a) Power extracted by the trailing foil for different kinematic configurations as a function of the Wake Phase parameter, $\Phi$ from \citet{Ribeiro2021}. These results scale nicely according to $\Phi$, showing that as the foil's kinematics bring it into alignment with the upstream wake structures ($\Phi\sim0^\circ$) its performance decreases. The cases highlighted by $\bigtriangleup$ markers and $\bigtriangledown$ markers demonstrate \textit{constructive} and \textit{destructive} interactions respectively, and are analyzed with PIV in Section \ref{sec:vortexFoilInteractions}. (b) Power extracted by the trailing foil for a range of $\theta_\mathrm{tr}$, all with $\Phi\approx180^\circ$, for each wake regime, with maximum $C_P^\mathrm{tr}$ values highlighted for each $\theta_\mathrm{tr}$. Data is scaled using the trailing foil's $\alpha_{T/4}^\mathrm{tr}$, where it is observed that maximum values of $C_P^\mathrm{tr}$ fall within a range of $0.32<\alpha_{T/4}^\mathrm{tr}<0.40$ rad, highlighted in green.}
    \label{fig:WakePhaseAndScaling}
\end{figure}

Figure~\ref{fig:WakePhaseAndScaling}a presents cases over a range of trailing foil heave, $h_\mathrm{tr}$, and foil separation, $S_x$, scaled using the \textit{wake phase parameter}, where we observe that cases with low-performance fall near $\Phi=0^\circ$ and those with high performance near $\Phi=\pm 180^\circ$.
Our results show an optimal wake phase of $\Phi \sim 100^\circ$ in the LEV+TEV regime, while \citet{Ribeiro2021} found a value of $\Phi=120^\circ$ in the same regime.
Overall, this indicates that for kinematics where the trailing foil is in alignment with the wake, we can expect poor performance in contrast to having the trailing foil avoid the wake, something also observed by \citet{Ribeiro2021}.
Additionally, as displayed by the predictive model of \citet{Ribeiro2024}, these results also display the sinusoidal-like trend of the power coefficient over the range of $\Phi$.

This, however, does not mean that wake-foil interactions always yield poor performance of the trailing foil.
As observed by \citet{Ashraf2011} and \citet{KinseyDumas2012}, some types of wake-foil interactions can lead to dramatically improved performance of the trailing foil.
As will be explained in Section \ref{sec:vortexFoilInteractions}, unexpectedly better performance can be obtained from the trailing foil due to favorable interactions with wake vortices.
These types of interactions were also captured by the results of \citet{Ribeiro2024}, where for the higher $\alpha_{T/4}$ values tested (0.4 rad and 0.49 rad) their model predicted vortex-foil interactions that led to significant increases in power output due to lift.
A similar conclusion can be obtained from looking at figure~\ref{fig:stackedPowerPhase}, as it is apparent that the large variation in performance in the LEV+TEV regime (where wake-foil interactions are presumed to be most influential) results in a system that is strongly affected by small changes in $\psi_{1-2}$.
From a practical perspective, it may be beneficial to operate within the LEV regime so that the trailing foil engages in weaker wake-foil interactions, yielding a more robust performance as well as slightly better system performance at the optimal operating kinematics.

Figure~\ref{fig:WakePhaseAndScaling}b shows cases with different trailing foil heave and pitch, $h_\mathrm{tr}$,  $\theta_\mathrm{tr}$,  but equal phase, $\Phi$.  The power is plotted as a function of the trailing foil $\alpha_{T/4}^\mathrm{tr}$. By noting that increasing $\theta_\mathrm{tr}$ shifts up the optimal value of $h_\mathrm{tr}$ for all wake regimes, it follows that larger $h_\mathrm{tr}$ at fixed $f^*$ increases $\dot{h}_\mathrm{tr}$ and therefore the power due to lift, $P_L=L \dot{h}_\mathrm{tr}$. However,  $\alpha_\mathrm{eff}$ will decrease (Eqn.~\ref{eq:aEff}) as $\dot{h}_\mathrm{tr}$ increases, acting to reduce the lift generated by the foil.
This is countered by raising $\theta_\mathrm{tr}$ to increase $\alpha_\mathrm{eff}$ and subsequently the lift, therefore obtaining higher power extraction.
Figure~\ref{fig:WakePhaseAndScaling}b demonstrates how the performance of the trailing foil should scale with $\alpha_{T/4}^\mathrm{tr}$, but only when subjected to the same upstream conditions.
These results show that there is an optimal range of $\alpha_{T/4}^\mathrm{tr}$ for all wake regimes at $0.32<\alpha_{T/4}^\mathrm{tr}<0.4$.
The findings presented by \citet{Ribeiro2024} indicate that further scaling of $C_P^\mathrm{tr}$ for each $\theta_\mathrm{tr}$ could be achieved by using the mean wake velocity upstream of the trailing foil.
Although this would work for cases where both foils have approximately the same heave amplitude, for cases where the trailing foil's heave is significantly larger this scaling would break down due to the foil's exposure to higher momentum flow (recall figure~\ref{fig:wakeVelocityAndVortices}).

\subsection{Optimal system performance.}
It is also interesting to recognize from all of these results (figures~\ref{fig:stackedPowerPitch} to \ref{fig:stackedPowerPhase}) that the maximum \textit{system power} extracted is obtained from the LEV wake regime ($\alpha_{T/4}^\mathrm{le} \sim 0.35$). In this case, even though the leading foil is not operating at its maximum capacity, the wake has more energy available for the trailing foil to capture and the system performance is maximized. 
Furthermore, the global optimum across all parameter combinations tested corresponds to the LEV regime with both foils operating with $f^*=0.11$.
Figure~\ref{fig:aT4vsEffV2} shows that a single foil's maximum energy harvesting efficiency can be achieved at $f^*>11$ when operating within the LEV+TEV regime.
In the tandem array, when the leading foil operated in the LEV regime, the $\alpha_{T/4}^\mathrm{le}$ value for the $f^*=0.11$ cases is slightly higher than for the $f^*=0.12$ cases ($0.35$ vs $0.33$ rad respectively), which accounts for the increase in performance of the leading foil.
Recall the performance bifurcation identified by \citet{RibeiroFranck2023} for higher $\alpha_{T/4}$ values, and that the lower branch in this bifurcation corresponds to $f^*\sim0.11$.
Cases with $f^*=0.11$ place the leading foil in that lower performance branch when operating in the LEV+TEV regime, explaining the notable drop in its performance with respect to cases with $f^*=0.12$.
Ultimately this results in significantly worse system performance within the LEV+TEV regime for the $f^*=0.11$ cases.

\subsection{Constructive and destructive vortex-foil interactions}
\label{sec:vortexFoilInteractions}
The system power performance results do not directly explain why some wake-foil interactions enhance trailing foil performance, while other kinematic choices degrade the trailing foil performance. \citet{Ashraf2011}, \citet{KinseyDumas2012}, \citet{Xu2016} and \citet{KarakasFenercioglu2017} identified some types of interactions that led to better or worse trailing foil performance. However,  in their tests both foils in the array shared the same heaving amplitude.
In the present study, due to the large differences in heave amplitude between the leading and trailing foil, the \textit{wake phase parameter} by itself does not make it immediately apparent what types of wake-foil interactions lead to high- and low-performance cases.

Detailed analysis on the three cases highlighted in figure~\ref{fig:WakePhaseAndScaling}a illustrates how certain kinematics, specifically $\psi_{1-2}$ and $h_\mathrm{tr}$, lead to improvements in performance.
We consider a \textit{constructive} interaction between the wake and the trailing foil as one that leads to improved trailing foil performance (up-facing $\bigtriangleup$ markers), while a \textit{destructive} interaction (down-facing $\bigtriangledown$ markers) is one resulting in decreased performance.
We evaluate cases in the LEV+TEV regime at the shortest spacing tested ($S_x=4c$) to amplify the effects of wake-foil interactions on the trailing foil, since the impact of the wake is stronger than in the lower $\alpha_{T/4}$ regimes, as implied by figure~\ref{fig:stackedPowerPhase}.

\begin{figure}
    \centering
    \includegraphics[width=\textwidth]{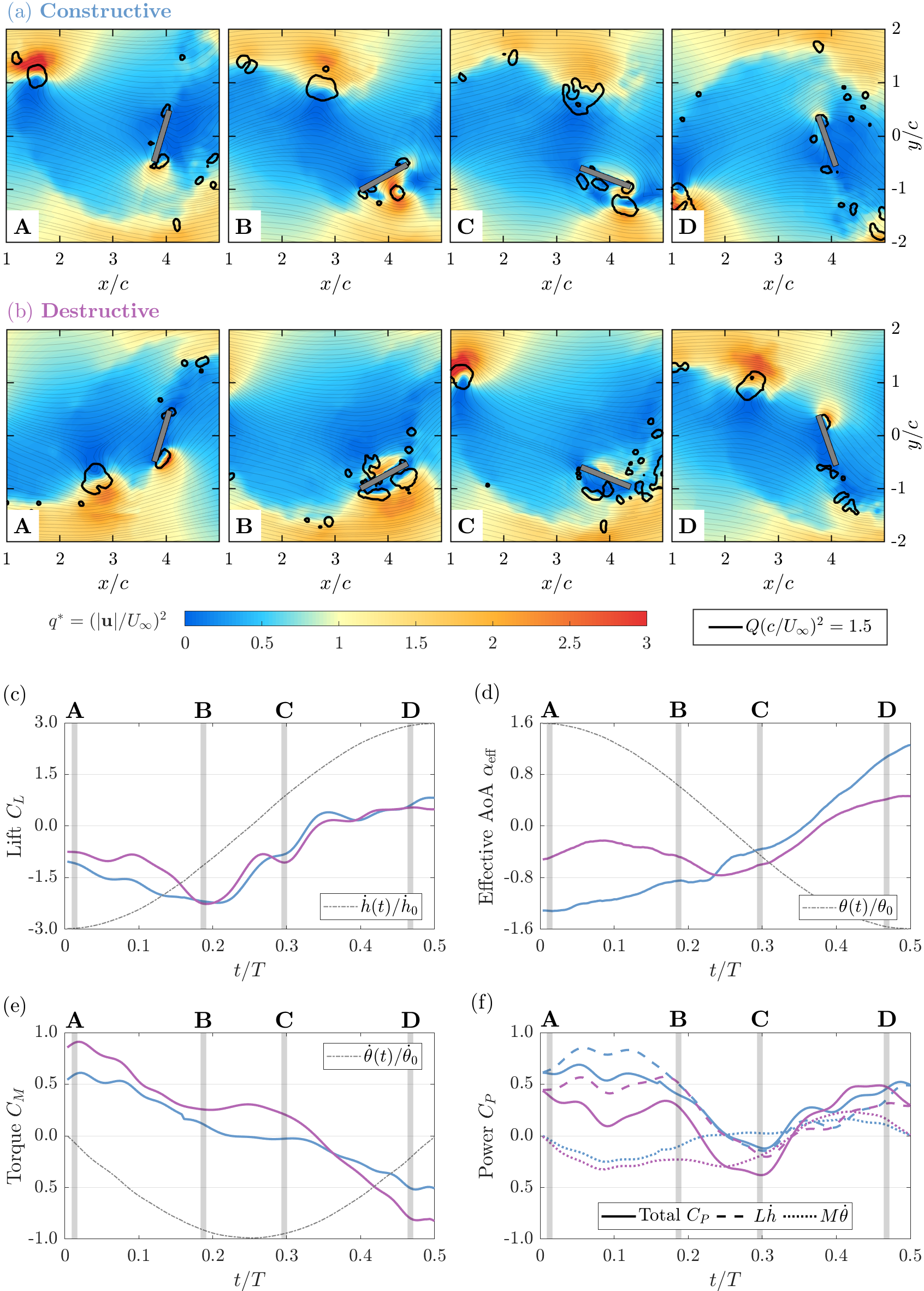}
    \caption{Snapshots of dynamic pressure $q^*=(|\textbf{u}|/U_\infty)^2$, with instantaneous streamlines of wake-foil interactions for $S_x=4c$, $\alpha_{T/4}=0.68$, $\theta_{\mathrm{tr}}=75^\circ$ and $h_\mathrm{tr}=0.8c$. Isolines of $Q(c/U_\infty)^2=1.5$ are shown to visualize primary vortices. Cases shown are for two values of inter-foil phase, $\psi_{1-2}=51^\circ$ for the constructive case in (a), and $\psi_{1-2}=180^\circ$ for the destructive case in (b). Also shown are non-dimensional (c) lift, (d) torque, (e) effective angle of attack, and (f) power over half of an oscillation cycle. (f) Also shows the power contributions from lift and torque.}
    \label{fig:WakeFoilInteractions8c}
\end{figure}

Two pairs of constructive and destructive interactions are compared in figures~\ref{fig:WakeFoilInteractions8c} and \ref{fig:WakeFoilInteractions14c}. Each figure shows snapshots from the PIV-measured flow field around the trailing foil taken at four instances in time (A-D) during half of the trailing foil's oscillation cycle. In both figures, the color shading represents the local dynamic pressure normalized by the free stream dynamic pressure, $q^*$:
\begin{equation}
    q^*=(|\mathbf{u}|/U_\infty)^2,
\end{equation}
where variations in $q^*$ are primarily due to the vortex-induced velocity. This metric allows us to determine when the trailing foil operates in more or less energetic regions of the wake, which we can relate to increases or decreases in the foil's energy harvesting performance. Instantaneous streamlines are shown to provide a sense of the direction of the flow. Thick black contour lines represent isolines of $Q$ (eq.~\ref{eq:Q-def}), non-dimensionalized appropriately: $Q(c/U_\infty)^2=1.5$ (note that $q^*$ and $Q$ represent different quantities).  These contour lines visualize the location of strong vortices in the wake.
Below the $q^*$ fields, the corresponding instantaneous effective angle of attack $\alpha_\mathrm{eff}$, lift coefficient $C_L$, moment coefficient $C_M$, and power coefficient $C_P$ curves of the trailing foil are shown for both the constructive (blue line) and destructive (purple line) cases.

The maximum and minimum $C_P^\mathrm{tr}$ cases at $h_\mathrm{tr}=0.8c$ are presented as \textit{constructive} and \textit{destructive} interactions in figures~\ref{fig:WakeFoilInteractions8c}a and \ref{fig:WakeFoilInteractions8c}b respectively.
In this configuration, both foils are exposed to the same flow window. The trailing foil operates only within the pure wake deficit region, which can be observed from the snapshots presented in figures~\ref{fig:WakeFoilInteractions8c}a-b, showing the foil positioned within the blue-shaded region in all snapshots. In the constructive case, shown in figure \ref{fig:WakeFoilInteractions8c}a, the trailing foil extracts an average power of $\overline{C^\mathrm{tr}_P}=0.299$ per cycle (blue $\bigtriangleup$ marker in figure~\ref{fig:WakePhaseAndScaling}a), while the destructive case in figure~\ref{fig:WakeFoilInteractions8c}b results in $\overline{C^\mathrm{tr}_P}=0.138$ (purple $\bigtriangledown$ marker in figure~\ref{fig:WakePhaseAndScaling}a).
These first two cases are of interest as they see the leading and trailing foils heaving with the same amplitude, and their global phase, $\Phi$, are approximately $180^\circ$ and $0^\circ$ respectively (see figure~\ref{fig:WakePhaseAndScaling}a).
Recalling that a wake phase of $\Phi=180^\circ$ indicates a case where the trailing foil is in anti-phase with the wake vortices, we can infer that given the choice of $\psi_{1-2}$ in the constructive case (and therefore the resulting $\Phi$), the trailing foil performs better due to the avoidance of vortices.
This can be observed in the PIV snapshots in figure~\ref{fig:WakeFoilInteractions8c}a, whereby following the primary LEV in the wake (large circle on the top left in snapshot A), we observe how the trailing foil avoids it as it travels downstream throughout the subsequent snapshots.
This contrasts the destructive case where, as observed in snapshots B and C in figure~\ref{fig:WakeFoilInteractions8c}b, the trailing foil collides head-on with the wake LEV.
Due to the lack of strong vortex-foil interactions, the constructive case would be considered a \textit{weak interaction} as described by \citet{KinseyDumas2012} and \citet{Xu2016}, while the direct collision with vortical structures in the destructive case makes it a \textit{strong wake interaction} case.

The difference in performance between these two cases is primarily due to the effective angle of attack the trailing foil experiences during its oscillation (figure~\ref{fig:WakeFoilInteractions8c}d). There is a large difference in $\alpha_\mathrm{eff}$ during instances A and B. By looking at the flow direction faced by the trailing foil in these two snapshots, we observe that the destructive case (figure~\ref{fig:WakeFoilInteractions8c}b, A-B) sees the trailing foil's $\alpha_\mathrm{eff}$ lowered because of the flow induced by the vortex shed from the leading foil (high $Q$ contour). This behavior was also observed by \citet{KinseyDumas2012} in their ``v4'' classification of wake foil interactions. In contrast, snapshots A and B of the constructive case (figure~\ref{fig:WakeFoilInteractions8c}a) show that the foil experiences a sustained high $\alpha_\mathrm{eff}$ (figure~\ref{fig:WakeFoilInteractions8c}d, A-B).  This 
results in higher lift (figure~\ref{fig:WakeFoilInteractions8c}c, A-B) at this initial part of the cycle.
This difference in lift has a significant impact on performance due to the high heaving velocity of the foil at this time that, because of its alignment with $C_L$, yields a high contribution to the power extracted (figure~\ref{fig:WakeFoilInteractions8c}f, A-B).

Turning attention the the stroke reversal (B-C in figure~\ref{fig:WakeFoilInteractions8c}), in the constructive case, we observe minimal torque and power from the LEV  as it moves over the foil towards the trailing edge (figure~\ref{fig:WakeFoilInteractions8c}a,e,f, B-C) . In contrast, the destructive case sees the attached LEV partially suppressed due to its collision with the wake LEV (figure~\ref{fig:WakeFoilInteractions8c}b, B-C) which results in a positive torque (figure~\ref{fig:WakeFoilInteractions8c}e, B-C) and a negative contribution to power (figure~\ref{fig:WakeFoilInteractions8c}f, B-C). Over the entire cycle, most of the energy extraction (figure~\ref{fig:WakeFoilInteractions8c}f) comes from lift production - in agreement with \citet{Kim2017} and \citet{Xu2016} - while most of the energy loss is due to the pitching torque during the stroke reversal (snapshots B-C).

Leading up to snapshot D both cases have similar performances before repeating the same behavior for the second half of the cycle.

\subsection{Effects of increased trailing heave amplitude}
\begin{figure}
    \centering
    \includegraphics[width=\textwidth]{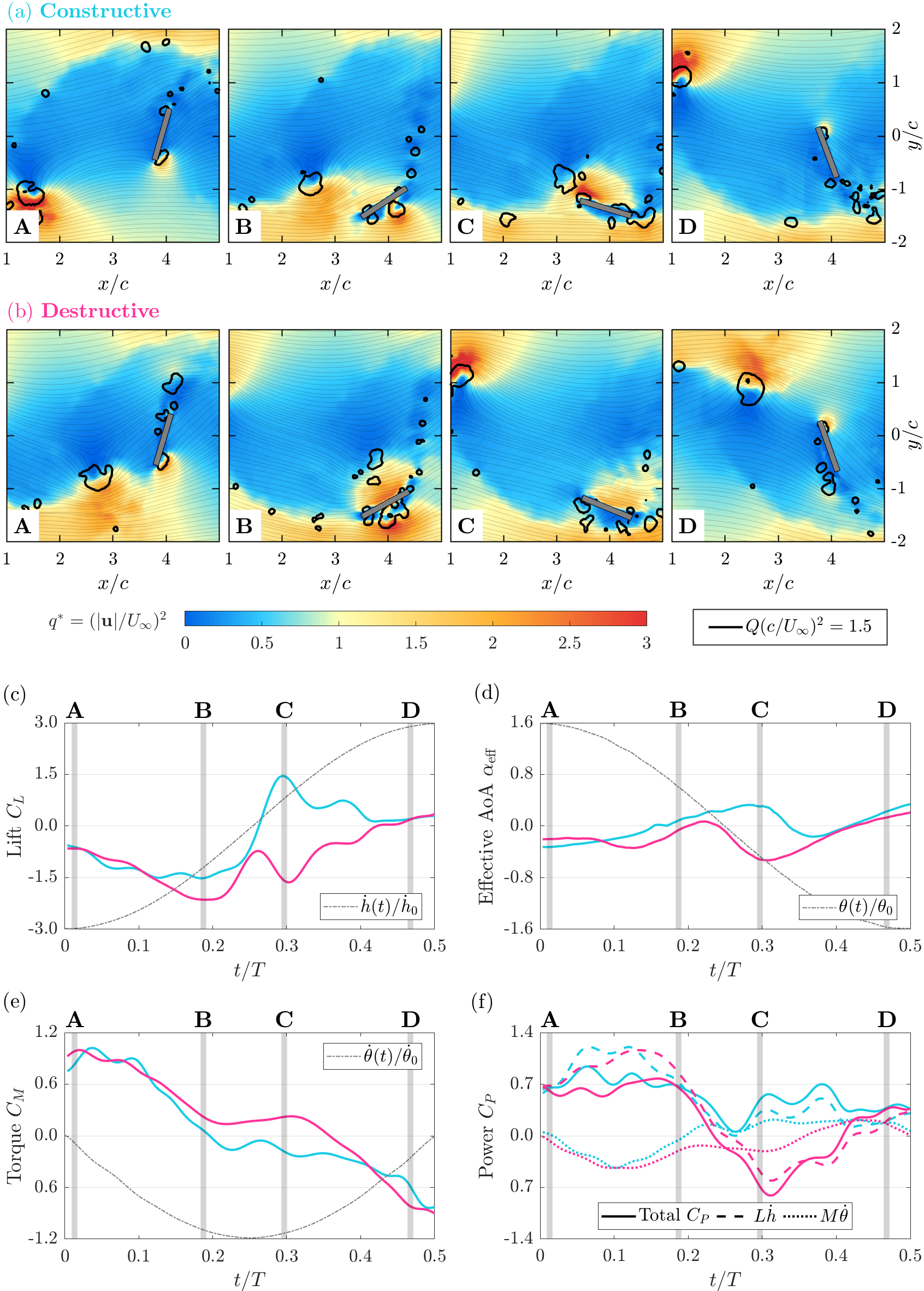}
    \caption{Snapshots of dynamic pressure $q^*=(|\textbf{u}|/U_\infty)^2$, with instantaneous streamlines of wake-foil interactions for $S_x=4c$, $\alpha_{T/4}=0.68$, $\theta_{\mathrm{tr}}=75^\circ$ and $h_\mathrm{tr}=1.4c$. Isolines of $Q(c/U_\infty)^2=1.5$ are shown to visualize primary vortices. Cases shown are for two values of inter-foil phase, $\psi_{1-2}=-110^\circ$ for the constructive case in (a), and $\psi_{1-2}=180^\circ$ for the destructive case in (b). Also shown are non-dimensional (c) lift, (d) torque, (e) effective angle of attack, and (f) power over half of an oscillation cycle. (f) Also shows the power contributions from lift and torque.}
    \label{fig:WakeFoilInteractions14c}
\end{figure}

The previous results, and all prior published results have focused on tandem foils that share the same heave amplitude.  However, figure~\ref{fig:stackedPowerHeave} demonstrates that increasing the trailing heave amplitude leads to increased system performance.  We can identify three reasons for this benefit. Firstly, the higher values of $h_{tr}$ sample higher momentum fluid (figure~\ref{fig:wakeVelocityAndVortices}a); secondly, larger heave amplitudes provide easier LEV avoidance, similar to that discussed in connection to figure~\ref{fig:WakeFoilInteractions8c}.  Lastly, the larger trailing heave amplitude does provide new mechanisms for constructive wake interactions. For example figure~\ref{fig:WakePhaseAndScaling}a shows a dramatic increase in system performance for a enlarged trailing heave due to a small shift in the wake phase, $\Phi$ (cyan and pink triangles).  These two cases are discussed in this section, and are illustrated in figure~\ref{fig:WakeFoilInteractions14c}.

In both the constructive and destructive cases (figure\ref{fig:WakeFoilInteractions14c}a,b), the trailing foil can reach the outer extent of the instantaneous wake region and take advantage of the accelerated velocity.
Note that both cases display direct collisions with wake vortices.   Using the wake interactions classification of \citet{KinseyDumas2012}, both examples shown in figure \ref{fig:WakeFoilInteractions14c} correspond to configurations that decrease instantaneous local dynamic pressure near the trailing foil, though the occurrence of this happens at slightly different times in the cycle, with the LEV reaching the trailing foil earlier in the destructive case.

It is immediately clear that the two cases have approximately equal performance at the points in cycle with high heave velocity, $\dot{h}$ (figure~\ref{fig:WakeFoilInteractions14c}A,D). This is likely due to flow being dominated by the attached LEV during this phase of the cycle. The  differences occur during the stroke reversal part of the oscillation (snapshots B-C).

In snapshot B of the constructive case (figure~\ref{fig:WakeFoilInteractions14c}a), as the foil reaches its maximum heave amplitude, it begins to shed its attached LEV and loses its lift, $C_L$ (figure~\ref{fig:WakeFoilInteractions14c}c, B). After this shedding occurs, the foil collides with the primary LEV in the wake (snapshot C) and gains $C_L$ in alignment with its heaving direction, generating positive power, $C_P,$ (figures~\ref{fig:WakeFoilInteractions14c}c,f C).
During the same part of the cycle, the destructive case sees the trailing foil go around the primary wake LEV and seemingly delay the shedding of its own attached LEV (figure~\ref{fig:WakeFoilInteractions14c}b, snapshot B).
Crucially, snapshot C in figure~\ref{fig:WakeFoilInteractions14c}b shows that the foil develops a second LEV, absent in the constructive case, although it can also be seen in the earlier configurations discussed (figures~\ref{fig:WakeFoilInteractions8c}a-b)
The effect of the second LEV is reflected in the lift, (figure~\ref{fig:WakeFoilInteractions14c}c C), which becomes more negative. This increase in $C_L$ results in negative $C_P$ being generated because the foil is heaving in the opposite direction at this time, (figure~\ref{fig:WakeFoilInteractions14c}f C). 

The $C_M$ performance is also affected by the presence or absence of the secondary LEV. Figure~\ref{fig:WakeFoilInteractions14c}e shows how, while the destructive case experiences positive $C_M$ during the stroke reversal (snapshots B and C), the constructive case $C_M$ becomes slightly negative. The effect is amplified by the high angular velocity at this time.

The result of the small difference in interfoil phase, $\psi_{1-2}$, (and consequently wake phase, $\Phi$) is that in the constructive case the foil experiences positive power throughout the cycle. Although this shows how the trailing foil can benefit greatly from vortex-foil interactions, the slight difference in $\psi_{1-2}$ between the constructive and destructive cases demonstrates how the trailing foil's performance is highly sensitive to small changes in kinematics when pursuing strong vortex-foil interactions. This echoes the result from \citet{Ma2019} who, in their semi-passive tandem foil system, noted that although favorable wake-foil interactions can yield significantly improved performance, it is more desirable to avoid negative wake-foil interactions than to pursue favorable ones.

\section{Conclusions}
The optimal kinematics and the role of wake-foil interactions on a tandem hydrofoil array's performance for energy harvesting were studied experimentally.  Distinct from previous such studies, we have chosen three operating regimes for the leading foil - the shear layer, LEV, and LEV+TEV regimes. For each of these regimes, we allow the trailing foil to have different kinematics (heave, pitch and phase) from the leading foil.
Special attention was paid to the effect that the wake structure prescribed by the leading foil has on the trailing foil. The role of the wake vortices produced by the leading foil on the trailing foil performance was also analyzed in detail.

The strongest effects from wake-foil interactions were observed in the LEV+TEV regime due to the primary vortices being stronger than the main LEV in the intermediate $\alpha_{T/4}$ regime.
Modifying the kinematics of the trailing foil, notably increasing the heaving amplitude, led to constructive wake-foil interactions previously not observed in the literature.  In this regime, the performance of the trailing foil is highly sensitive to the phase, $\psi_\mathrm{1-2}$,  between the leading and trailing foils, and we find that dramatically improved or reduced performance can result due to constructive or destructive wake vortex-foil interactions.

The highest system performance observed in the parameter space explored was when the leading foil operated within the LEV regime, different from the single-foil optimum, which is in the LEV+TEV regime.
In the LEV regime, the trailing foil does not see its performance affected as dramatically by wake-foil interactions, and the wake-deficit encountered is not as strong as in the LEV+TEV regime.
This results in the interesting conclusion that for the tandem array to perform optimally, neither foil should operate at its single-foil optimal kinematics.
It was also found that although the primary vortices in the LEV and LEV+TEV regimes initially follow different paths after being shed by the leading foil, their trajectories converge as they travel downstream.
This results in the primary vortex approaching the trailing foil with the same $y$-location across wake regimes.
As observed by previous literature the inter-foil phase directly affects the timing of wake-foil interactions, but we additionally find that the heaving amplitude of the trailing foil has a significant impact on the type of interactions that occurred, and consequently the harvesting performance.

As has been previously noted, and is studied in some detail here, when the leading and trailing heave amplitudes are matched, constructive interactions are observed for cases that avoid wake structures, while destructive interactions occur when the trailing foil collides with the leading foil LEV.  However, a second kind of constructive interaction is also identified when the trailing foil has heave amplitude larger than the leading foil. In this case, the leading vortex does interact with the trailing foil, but acts to \emph{enhance} its energy extraction.

These results reinforce the appeal of oscillating flow turbines. While the power coefficient of the individual turbines is lower than, for example, a single horizontal axes turbine, due to the ability to locate tandem turbines in close proximity, the system power density is far higher than that of a conventional turbine array. This operational advantage, combined with other features of OFTs, such as their insensitivity to specific blade design \citep{Kim2017} may lead to greater adoption in renewable energy harvesting applications.

\section*{Acknowledgements}

This work was supported by the National Science Foundation, Grant 1921359.  The authors are most grateful to Jen Franck and Bernardo Ribeiro for many fruitful discussions.

\bibliographystyle{jfm}

\bibliography{references}

\end{document}